# ANALYSIS AND SYNTHESIS OF THE SNS SUPERCONDUCTING RF CONTROL SYSTEM

Y.M. Wang, S.I. Kwon, and A.H. Regan, LANL, Los Alamos, NM 87545, USA

*Abstract*

The RF system for the SNS superconducting linac consists of a superconducting cavity, a klystron, and a low-level RF (LLRF) control system. For a proton linac like SNS, the field in each individual cavity needs to be controlled to meet the overall system requirements. The purpose of the LLRF control system is to maintain the RF cavity field to a desired magnitude and phase by controlling the klystron driver signal. The Lorentz force detuning causes the shift of the resonant frequency during the normal operation in the order of a few hundreds hertz. In order to compensate the Lorentz force detuning effects, the cavity is pre-tuned into the middle of the expected frequency shift caused by the Lorentz force detuning. Meanwhile, to reduce the overshoot in the transient response, a feed-forward algorithm, a linear parameter varying gain scheduling (LPV-GS) controller, is proposed to get away a repetitive noised caused by the pulsed operation as well as the Lorentz force detuning effects.

## 1 INTRODUCTION

To analyse the performance of the RF control system for the SNS superconducting linac, a MATLAB model is created for each functional blocks, which includes the superconducting cavity model, klystron model, PID feedback controller, and a feed-forward controller[1]. An equivalent resonant circuit couple with a coupling transformer is used for the superconducting cavity model in which the Lorentz force detuning of the cavity resonance frequency is included. The klystron is modelled as a cascade of a pass filter, determined by the bandwidth of the klystron, and a phase-magnitude saturation curve, which represents the saturation characteristics of the klystron. The phase-magnitude saturation curve is obtained from the measurement and is further analysed using the curve fitting to generate the final model. The main feedback controller is a PI controller for an easy implementation and robustness concern. In order to implement the RF control system in a full digital control system, the latency analysis is needed to satisfy the performance requirement of the system. Finally, with the results obtained from the numerical simulation and the performance requirements, a full digital control system for the LLRF system is proposed. In this system, a combined CPLD and DSP technology is used to cope with different requirements. The CPLD is applied to the critical path in which the time delay needs to be minimized. While the DSP is used to perform the complex linear parameter varying gain scheduling (LPV-GS) control which requires the computation power but needs only be fed to the control signal in the next pulse.

## 2 SYSTEM MODELLING AND CONTROL ALGORITHMS

### 2.1 Superconducting Cavity Model

The state space equation of the superconducting model is given by

$$\dot{x} = A(\Delta\omega_L)x + Bu + B_I I \quad (1)$$
$$y = Cx$$

where,

$$A(\Delta\omega_L) = \begin{bmatrix} -\dfrac{1}{\tau_L} & -(\Delta\omega_m + \Delta\omega_L) \\ (\Delta\omega_m + \Delta\omega_L) & -\dfrac{1}{\tau_L} \end{bmatrix}$$

the dynamics of the Lorentz force detuning satisfies the following equation

$$\dot{\Delta\omega}_L = -\frac{1}{\tau_m}\Delta\omega_L - \frac{2\pi}{\tau_m}\overline{K}V_I^2 - \frac{2\pi}{\tau_m}\overline{K}V_Q^2 \quad (2)$$

where, $\Delta\omega_m$ is the synchronous phase detuning frequency, $\Delta\omega_L$ is the Lorentz force detuning frequency, $\tau_L$ is the loaded cavity damping constant, $K$ is the Lorentz force detuning constant, $x = \begin{bmatrix} V_I & V_Q \end{bmatrix}$ is the cavity field in I/Q components, whereas, the system matrices B, $B_I$, and C are given in [1].

In the model, the Lorentz force detuning frequency appears on in the system matrix A and all other system matrices are constant. In observing Equation (2), the Lorentz force detuning is a nonlinear function of the cavity field, which renders the system equation (1) a nonlinear equation of the cavity field.

## 2.2 Linear Parameter Varying Gain Scheduling Controller (LPV GS)

The principles of the linear parameter varying gain scheduling can be explained as the followings. First, due to the nonlinearity of the system equation, which comes from both the saturation characteristic of the klystron and the nature of the Lorentz force detuning effect, the maximum performance of the RF control system can only be achieved by implementing a variable gain-profile based on the equilibrium point at which the system operates. Secondly, at the equilibrium point, the system needs to be linearized for solving the system equation (1). Finally, both the feedback controller and the feed-forward controller need to be implemented to suppress the repetitive noise due to the pulsed operation and a known effect of the Lorentz force detuning effect.

The equilibrium manifold of a linear parameter varying system is given by

$$\dot{x} = A(\rho)x + B(\rho)u + E(\rho)w \qquad (3)$$
$$y = C(\rho)x .$$

The above equations are a linearized version of the system equation (1) at a specific operation point given by $\rho$. Let $y_r$ be the desired trajectory to be followed by the system output $y$. Then, the parameterised equilibrium manifold of the system is defined by the solution of the algebraic equation given

$$\begin{bmatrix} 0 \\ y_r \end{bmatrix} - \begin{bmatrix} E(\rho)w \\ 0 \end{bmatrix} = \begin{bmatrix} A(\rho) & B(\rho) \\ C(\rho) & 0 \end{bmatrix} \begin{bmatrix} x_e \\ u_e \end{bmatrix} \qquad (4)$$

Now we consider the open loop system as given in (1) and the Lorentz force detuning as given in (2). First, let

$$V = \begin{bmatrix} v_I & v_Q \end{bmatrix}^T$$

be the desired output trajectory to be tracked by the cavity field I and Q. Then, the equilibrium manifold $(x_e, u_e)$ of the open loop system as given in (1) is the solution of the following algebraic matrix equation.

$$\begin{bmatrix} 0 \\ -- \\ V \end{bmatrix} = \begin{bmatrix} -\frac{1}{\tau_L} & -(\Delta\omega_m + \Delta\omega_L) & | & \frac{2}{Z_o}c_1 & -\frac{2}{Z_o}c_3 \\ (\Delta\omega_m + \Delta\omega_L) & -\frac{1}{\tau_L} & | & \frac{2}{Z_o}c_3 & \frac{2}{Z_o}c_1 \\ ---------- & & | & ---------- \\ 1 & 0 & | & 0 & 0 \\ 0 & 1 & | & 0 & 0 \end{bmatrix} \begin{bmatrix} x_e \\ u_e \end{bmatrix} + \begin{bmatrix} -2c_1\zeta & 2c_3\zeta \\ -2c_3\zeta & -2c_1\zeta \\ 0 & 0 \\ 0 & 0 \end{bmatrix} I \qquad (5)$$

Solving Equation (5), we obtain

$$x_e = V \qquad (6)$$

$$u_e = -\frac{Z_o}{2(c_1^2 + c_3^2)} \begin{bmatrix} c_1 & c_3 \\ -c_3 & c_1 \end{bmatrix} \left( \begin{bmatrix} -\frac{1}{\tau_L} & -(\Delta\omega_m + \Delta\omega_L) \\ (\Delta\omega_m + \Delta\omega_L) & -\frac{1}{\tau_L} \end{bmatrix} V + \begin{bmatrix} -2c_1\zeta & 2c_3\zeta \\ -2c_3\zeta & -2c_1\zeta \end{bmatrix} I \right) \qquad (7)$$

Note that the equilibrium manifold $(x_e, u_e)$ is parameterized by not only the desired trajectory $V$, the Lorentz force detuning $\Delta\omega_L$ but also the beam current I.

From (2), the Lorentz force detuning on the equilibrium manifold is

$$\Delta\omega_{Le} = -2\pi \overline{K} x_{1e}^2 - 2\pi \overline{K} x_{2e}^2 . \qquad (8)$$

Using the equilibrium points obtained from (6) and (7), we can design a linear parameter varying gain-scheduling controller as

$$u = u_e + F(\Delta\omega_L, I, V)(x - x_e) \qquad (9)$$

In the controller (9), $F(\Delta\omega_L, I, V)$ is the parameter varying feedback gain matrix such that the closed loop system matrix

$$A_{cl}(\Delta\omega_L) = A(\Delta\omega_L) + BF(\Delta\omega_L, I, V) \qquad (10)$$

is stable.

There are many design techniques for $F(\Delta\omega_L, I, V)$. A $H_\infty$ controller-based parametric varying controller and a velocity-based gain-scheduling controller are two of them. In addition, we can design a constant feedback gain matrix $F$ such that for all variations of $\Delta\omega_L$, $V$, and $I$ within given bounded sets, the closed loop system matrix (10) is stable. An eigenstructure control design technique can be applied. Let the constant stable matrix $A_r$ be the desired closed loop system matrix. Then, the feedback controller gain matrix $F(\Delta\omega_L, I, V)$ is determined by solving

$$A_r(\Delta\omega_L) = A(\Delta\omega_L) + BF(\Delta\omega_L, I, V) \qquad (10)$$

The solution of Equation (11) is

$$F(\Delta\omega_L, I, V) = B^{-1}(A_r - A(\Delta\omega_L)) \qquad (10)$$

Assume that the desired closed loop system matrix is a diagonal matrix given by

$$A_r = \begin{bmatrix} a_{r1} & 0 \\ 0 & a_{r2} \end{bmatrix}.$$

Then,

$$F(\Delta\omega_L, I, V) = \frac{Z_o}{2} \frac{1}{c_1^2 + c_2^2} \begin{bmatrix} F_{11} & F_{12} \\ F_{21} & F_{22} \end{bmatrix} \qquad (13)$$

where,

$$F_{11} = c_1(a_{r1} + \frac{1}{\tau_L}) - c_3(\Delta\omega_m + \Delta\omega_L)$$

$$F_{12} = c_3(a_{r2} + \frac{1}{\tau_L}) + c_1(\Delta\omega_m + \Delta\omega_L)$$

$$F_{21} = -c_3(a_{r1} + \frac{1}{\tau_L}) - c_1(\Delta\omega_m + \Delta\omega_L)$$

$$F_{22} = c_1(a_{r2} + \frac{1}{\tau_L}) - c_3(\Delta\omega_m + \Delta\omega_L)$$

The controller as given in (9) together with (6), (7), and (12) is a parametrically dependent controller where the Lorentz force detuning $\Delta\omega_L$, beam current $I$, and the desired trajectory $V$ are parameters defining the controller [1].

## 3 SIMULATION RESULTS AND CONCLUSIONS

Figure 1 is the block diagram of the RF control system. As we can see that the fast signal path is the implemented using the CPLD while the error feed-forward is implemented using the DSP. The total frequency response of the system is given is Figure 2 illustrates the effect of the Lorentz force detuning on the pole locations.

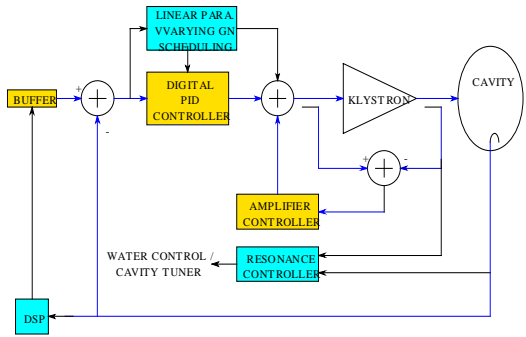

Figure 1. The block diagram of the RF control system.

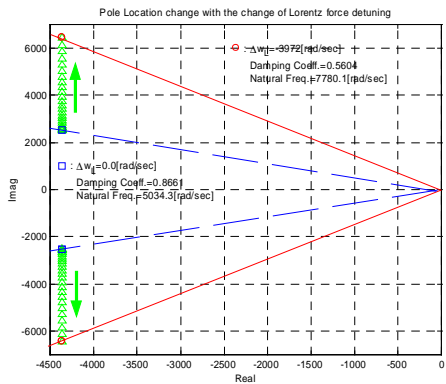

Figure 2. Root loci of the characteristic equation

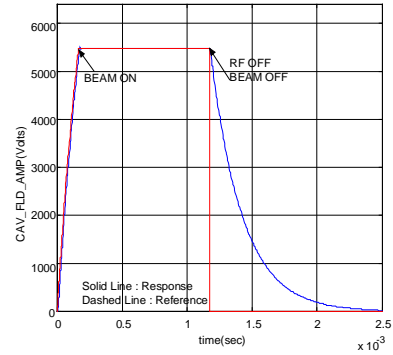

Figure 3. Field amplitude response for a closed-loop system with a LPV-GS controller.

The system performance is given in Figure 3 in which the steady state value is within the error limit. In Figure 4, the performance of the feed-forward control is represented in a way so that the reduction of the repetitive noise due to the beam pulse can be observed.

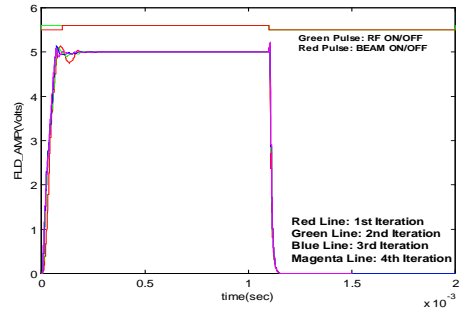

Figure 4. Pulse to pulse responses of the cavity field with a LPV-GS controller.

From the analysis and the simulation results obtained from our modelling, it is obviously that the performance requirements have been achieved with a full digital control system in which the latency of the digital system has been take into account in the modelling. However, in the real operation, other problems may arise, such as the effect of the microphonics. The performance of the proposed RF control system in the real operation will be reported when the data is available.